\documentclass[aps,preprint,onecolumn]{revtex4}
\usepackage{amsfonts}
\usepackage{amsmath}
\usepackage{amssymb}
\usepackage{graphicx}

\input{tcilatex}
\begin{document}

\title{An alternative realization of spontaneous emission cancellation via
Field Generated Coherence (FGC)}
\author{Fazal Ghafoor}
\affiliation{Department of Physics, COMSATS Institute of Information Technology,
Islamabad, Pakistan}

\begin{abstract}
In contrast to the traditional Spontaneous Generated Coherence
(SGC), Field Generated Coherence (FGC)-based atomic scheme is
presented for spontaneous emission cancellation. It is easy to
achieve externally controllable experimental trapping condition in
this 4-field-driven 5-level atomic system. Consequently, due to
the FGC the decay from the central dressed
\emph{bare-energy-state} of the set of upper three closely spaced
hyperfine decaying states of Sodium D2 line is completely
cancelled under the trapping
condition, exhibiting a novel phenomenon of a \textit{dark bare-energy-state}%
. Extending to an atomic system of simple probability loss, based
on Sodium D1 line, the bright atom can also be darkened under its
trapping condition, representing another experimentally viable,
novel and interesting phenomenon.
\end{abstract}

\date[Date text]{date}
\maketitle

The interaction of atoms or molecules with the environmental modes leads to
spontaneous emission in atomic systems. The simplest example is the free
space where atomic coherence and quantum interference are the basic
mechanisms for cancellation \cite{Agarwal,Alzetta,Gray} of spontaneous
emission, a basic phenomenon not questionable regarding its utilities \cite%
{Garraway,Zhu,Kochar,Scully}. On the basis of its mechanisms we
can divide it into two main categories. The first is spontaneous
emission generated coherence (SGC) where the decay processes
generate coherence among themselves to cancel spontaneous emission
\cite{Zhu-Scully,Paspalakis}. The second mechanism depends on the
driving fields itself where one coherence induces the others. This
is intuitively the simplest mechanism which may be easily realized
in a laboratory, and is the subject of this letter. We introduce a
system based on Fields Generated Coherence (FGC), a collective
coherence effect of amplitudes and phases of the driving fields on
the spontaneous emission processes. Spontaneous emission can be
cancelled under a field-dependent trapping condition along with
the other atomic population transfer effects among the three
decaying dressed \textit{bare-energy-states} making the central
one completely dark, an unexpected but viably novel phenomenon.
Remarkably, the trapping condition achieved for this system is
externally controllable and easy to implement experimentally.
Furthermore, the same concept can be extended to an atomic system
of simple probability loss adjustable with a recent experiment to
darken the brightened atom. This is an amazing and interesting
phenomenon leading to the trapping of all the population in the
unique excited decaying \textit{bare-energy-state}. Generally, all
population in this one-atom quantum system may be transferred into
a unique, extremely slowly decaying dressed state, allowing
effective storage and manipulation of atomic population like in
Ref. \cite{2-atom} but with the additional darkened
\emph{bare-energy-state}. It is worthwhile to note the confusion
of the terminology in literature between the control and cancellation \cite%
{Paspalakis} of spontaneous emission which needs clarification
\cite{F-Ghafoor}.

Prior to discussing the physics of the FGC regarding the spontaneous
emission cancellation in our proposed scheme, let us recall briefly some
pioneer works carried out in the area of SGC and its complications. For
example, Zhu and Scully \cite{Zhu-Scully} observed spectral line elimination
associated with dressed state in a four-level atomic system, arising due to
quantum interference effect between the upper decaying non-degenerate two
levels to the same lower level. Further, Paspalakis and Knight proposed a
phase control scheme in a four-level atom driven by two lasers of the same
frequency \cite{Paspalakis}, where the relative phase of the two lasers was
used to get extreme line-width narrowing, partial control of all the three
dressed-state and total cancellation of a dressed-state in the spontaneous
emission spectrum. The beautiful physics of these processes is also
explained in Ref. \cite{Hwang-Zhu} by using dressed state-vector approach.
However, all the spontaneous emission cancellation schemes have one common
origin, that is, the decay processes from two closely spaced atomic levels
to a third level with a condition of parallel dipole moments. Two closely
spaced levels can hardly be created by mixing two-parity levels due to
static electric field. For example, a separation of even 40$\gamma $ for $%
\left\vert 2s\right\rangle $\ and $\left\vert 2p\right\rangle $ states of
hydrogen atom \cite{Hakuta} could not utilized for successful demonstration
of the processes. It is hard to satisfy simultaneously the rigorous
condition for the spontaneously generated coherence of nearly degenerate
levels and the parallel dipole moments. Consequently, some experiments have
been performed (see Ref. \cite{Xia}) but with a doubt, as commented upon in
Ref. \cite{Li}. Ultimately, a scheme based on orthogonal dipole moments \cite%
{F-Ghafoor,Ghafoor00} may lead to experimentally more realistic system if it
qualifies for spontaneous emission cancellation under a viably novel
phenomenon. In the following this approach is developed.

\begin{figure}[t]
\centering
\includegraphics[width=5.5in]{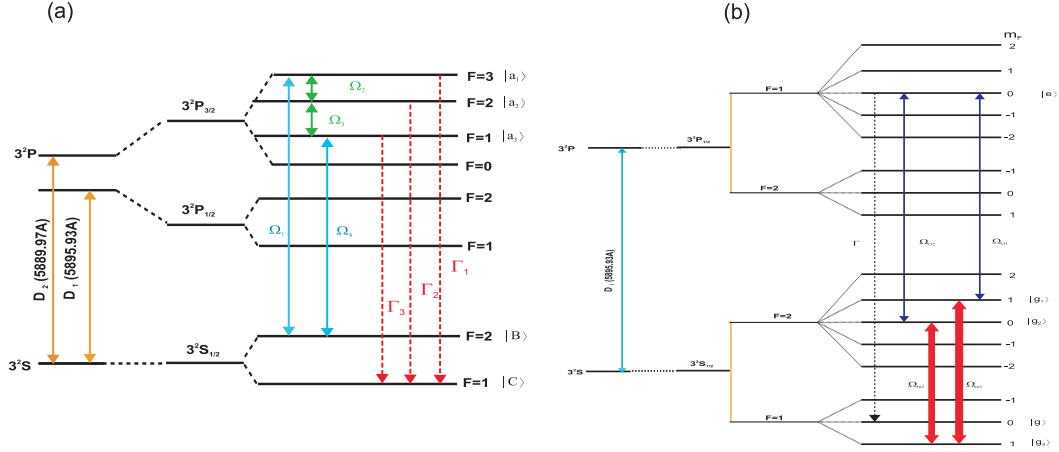}
\caption{ Schematics (a) Hyperfine-structured Sodium D2 line. (b):
Zeeman-structured Sodium D1 line} \label{figure1}
\end{figure}

Consider a 5-level atomic system based on the hyperfine-structured
Sodium D2 line (3S$_{1/2}^{2}\Longleftrightarrow $3P$_{3/2}^{2}$)
[see Fig. 1(a)]. Two
pairs of 3$^{2}$P$_{3/2},$F=1 ($\left\vert a_{3}\right\rangle $), 3$^{2}$P$%
_{3/2},$F=2 ($\left\vert a_{2}\right\rangle $), 3$^{2}$P$_{3/2},$F=3 ($%
\left\vert a_{1}\right\rangle $) from the excited quadruplet are
driven by microwave fields to have the Rabi-frequencies $\Omega
_{1}$ and $\Omega _{2}$ respectively. The ground state
3$^{2}$S$_{1/2},$F=2 ($\left\vert B\right\rangle $) is coupled
with the excited states $\left\vert a_{1}\right\rangle $ and
$\left\vert a_{3}\right\rangle $ via two coherent fields to have
Rabi frequencies $\Omega _{3}$ and $\Omega _{4\text{ }}$
respectively. The selected three closely spaced states decay to
another ground state 3$^{2}$S$_{1/2},$F=1 ($\left\vert
C\right\rangle $) (allowed transitions) by the vacuum field modes
couplings. Now we have to measure the spectrum in steady state
limit. Furthermore, the experiment of Xia \emph{et al.} \cite{Xia}
using sodium dimer\textit{\ }may also be adjusted with this set-up
using the naturally existing series of coupled excited energy
states arising due to mixing of the triplet and singlet g-parity
Rydberg states by the spin-orbit coupling known as occasional
perturbation \cite{Herz,Wang}.

Using the Weisskofp-Wigner theory the equations of motion
for the probability amplitudes of the atomic system are obtained as%
\begin{align}
\overset{\cdot }{A}_{1}\left( t\right) & =-i\Omega _{2}e^{i\Delta
_{2}t}A_{2}\left( t\right) -i\Omega _{1}e^{i\Delta _{1}t}B\left( t\right) -%
\frac{\Gamma _{1}}{2}A_{1}\left( t\right)  \notag \\
& -p_{1}\frac{\sqrt{\Gamma _{1}\Gamma _{2}}}{2}e^{-i\omega
_{12}t}A_{2}\left( t\right) -p_{2}\frac{\sqrt{\Gamma _{1}\Gamma _{3}}}{2}%
e^{-i\omega _{13}t}A_{3}\left( t\right) , \\
\overset{\cdot }{A}_{2}\left( t\right) & =-i\Omega _{2}^{\ast }e^{-i\Delta
_{2}t}A_{1}\left( t\right) -i\Omega _{3}e^{i\Delta _{3}t}A_{3}\left(
t\right) -\frac{\Gamma _{2}}{2}A_{2}\left( t\right)  \notag \\
& -p_{1}\frac{\sqrt{\Gamma _{1}\Gamma _{2}}}{2}e^{-i\omega
_{12}t}A_{1}\left( t\right) -p_{3}\frac{\sqrt{\Gamma _{2}\Gamma _{3}}}{2}%
e^{-i\omega _{23}t}A_{3}\left( t\right) , \\
\overset{\cdot }{A}_{3}\left( t\right) & =-i\Omega _{3}^{\ast }e^{-i\Delta
_{3}t}A_{2}\left( t\right) -i\Omega _{2}e^{i\Delta _{4}t}B\left( t\right) -%
\frac{\Gamma _{3}}{2}A_{3}\left( t\right)  \notag \\
& -p_{2}\frac{\sqrt{\Gamma _{1}\Gamma _{3}}}{2}e^{-i\omega
_{13}t}A_{1}\left( t\right) -p_{3}\frac{\sqrt{\Gamma _{2}\Gamma _{3}}}{2}%
e^{-i\omega _{23}t}A_{2}\left( t\right) , \\
\overset{\cdot }{B}\left( t\right) & =-i\Omega _{1}^{\ast }e^{-i\Delta
_{1}t}A_{1}\left( t\right) -i\Omega _{4}^{\ast }e^{-i\Delta
_{4}t}A_{3}\left( t\right) , \\
\overset{\cdot }{C_{k}}\left( t\right) & =-ig_{\mathbf{k}}^{\ast
(1)}e^{-i\delta _{1}t}A_{1}\left( t\right) -ig_{\mathbf{k}}^{\ast
(2)}e^{-i\delta _{2}t}A_{2}\left( t\right) -ig_{\mathbf{k}}^{\ast
(3)}e^{-i\delta _{3}t}A_{3}\left( t\right) ,
\end{align}%
where $\Gamma _{j}$ $\left( j=1,2,3\right) $ are the radiative decay rates
from the upper three levels to the ground level, respectively. Further $%
\Delta _{1}=\omega _{a_{1}}-\omega _{b}-\upsilon _{1},$ $\Delta _{2}=\omega
_{a_{1}}-\omega _{a_{2}}-\upsilon _{2},$ $\Delta _{3}=$ $\omega
_{a_{2}}-\omega _{a_{3}}-\upsilon _{3},$ and $\Delta _{4}=\omega
_{a_{3}}-\omega _{b}-\upsilon _{4}$ are the driving fields detunings, $%
\delta _{1}=\omega _{a_{1}}-\omega _{c}-\upsilon _{\mathbf{k}},$ $\delta
_{2}=\omega _{a_{2}}-\omega _{c}-\upsilon _{\mathbf{k}},$ and $\delta
_{3}=\omega _{a_{3}}-\omega _{c}-\upsilon _{\mathbf{k}},$ are the vacuum
fields detunings while $g_{\mathbf{k}}^{\left( 1\right) },$ $g_{\mathbf{k}%
}^{\left( 2\right) }$ and $g_{\mathbf{k}}^{\left( 3\right) }$ are
the vacuum field coupling constants, respectively. However, the
alignments  $p_{i}=\mathbf{r}%
_{a_{i},c}.\mathbf{r}_{a_{j},c}/r_{a_{i},c}r_{a_{j},c}$ $\vee $
$(i=1-3$ $\&$ $j=i)$ of the matrix elements among the three dipole
moments are neglecting under the approximations $\omega _{12},$
$\omega _{23}\gg \Gamma _{1,2,3}$ \cite{F-Ghafoor,Ghafoor00}.

Using Laplace transforms and the final value theorem along with
choosing the detuning parameters $\delta _{i}$ $\left(
i=1-3\right) =$ $\delta +\omega _{12},$ $\delta ,$ $\delta -\omega
_{12}$ $\vee $ $\delta =\omega _{a_{2}c}-\nu _{k}$ and $\omega
_{12}\approx \omega _{23},$ the steady state
probability amplitudes are given by%
\begin{align}
\mathcal{A}_{1}\left( \delta +\omega _{12}\right) & =\frac{\mathcal{A}%
_{1}\left( 0\right) }{\mathcal{D}\left( \delta +\omega _{12}\right) }%
[i\left( \delta +\omega _{12}\right) [i\left( \delta +\omega _{12}\right) +%
\frac{\Gamma _{2}}{2}][i\left( \delta +\omega _{12}\right) +\frac{\Gamma _{3}%
}{2}]  \notag \\
& +[i\left( \delta +\omega _{12}\right) +\frac{\Gamma _{2}}{2}]\left\vert
\Omega _{4}\right\vert ^{2}+i\left( \delta +\omega _{12}\right) \left\vert
\Omega _{3}\right\vert ^{2}]  \notag \\
& +\frac{\mathcal{A}_{2}\left( 0\right) }{\mathcal{D}\left( \delta +\omega
_{12}\right) }\left[ \left( \delta +\omega _{12}\right) \left[ i\left(
\delta +\omega _{12}\right) +\frac{\Gamma _{3}}{2}\right] \Omega
_{2}-i\Omega _{2}\left\vert \Omega _{4}\right\vert ^{2}+i\Omega _{1}\Omega
_{3}^{\ast }\Omega _{4}^{\ast }\right]  \notag \\
& -\frac{\mathcal{A}_{3}\left( 0\right) }{\mathcal{D}\left( \delta +\omega
_{12}\right) }\left[ i\left( \delta +\omega _{12}\right) \Omega _{2}\Omega
_{3}+\left[ i\left( \delta +\omega _{12}\right) +\frac{\Gamma _{2}}{2}\right]
\Omega _{1}\Omega _{4}^{\ast }\right]  \notag \\
& +i\frac{\mathcal{B}\left( 0\right) }{\mathcal{D}\left( \delta +\omega
_{12}\right) }\left[ \Omega _{2}\Omega _{3}\Omega _{4}-\left[ i\left( \delta
+\omega _{12}\right) +\frac{\Gamma _{3}}{2}\right] \left[ i\left( \delta
+\omega _{12}\right) +\frac{\Gamma _{2}}{2}\right] \Omega _{1}-\Omega
_{1}\left\vert \Omega _{3}\right\vert ^{2}\right] ,
\end{align}%
\begin{align}
\mathcal{A}_{2}\left( \delta \right) & =i\frac{\mathcal{A}_{1}\left(
0\right) }{\mathcal{D}\left( \delta \right) }\left[ -i\delta \left( i\delta +%
\frac{\Gamma _{3}}{2}\right) \Omega _{2}^{\ast }+\Omega _{2}^{\ast
}\left\vert \Omega _{4}\right\vert ^{2}-\Omega _{1}^{\ast }\Omega _{3}\Omega
_{4}\right]  \notag \\
& -\frac{\mathcal{A}_{2}\left( 0\right) }{\mathcal{D}\left( \delta \right) }%
\left[ i\delta \left( i\delta +\frac{\Gamma _{1}}{2}\right) \left( i\delta +%
\frac{\Gamma _{3}}{2}\right) +\left( i\delta +\frac{\Gamma _{1}}{2}\right)
\left\vert \Omega _{4}\right\vert ^{2}+\left( i\delta +\frac{\Gamma _{3}}{2}%
\right) \left\vert \Omega _{1}\right\vert ^{2}\right]  \notag \\
& +i\frac{\mathcal{A}_{3}\left( 0\right) }{\mathcal{D}\left( \delta \right) }%
\left[ -\Omega _{1}\Omega _{2}^{\ast }\Omega _{4}^{\ast }-i\delta \left(
i\delta +\frac{\Gamma _{1}}{2}\right) \Omega _{3}-\Omega _{3}\left\vert
\Omega _{1}\right\vert ^{2}\right]  \notag \\
& -\frac{\mathcal{B}\left( 0\right) }{\mathcal{D}\left( \delta \right) }%
\left[ \left( i\delta +\frac{\Gamma _{1}}{2}\right) \Omega _{3}\Omega
_{4}+\left( i\delta +\frac{\Gamma _{3}}{2}\right) \Omega _{1}\Omega
_{2}^{\ast }\right] ,
\end{align}%
and%
\begin{align}
\mathcal{A}_{3}\left( \delta -\omega _{12}\right) & =\frac{\mathcal{A}%
_{1}\left( 0\right) }{\mathcal{D}\left( \delta -\omega _{12}\right) }\left[
i\left( \delta -\omega _{12}\right) \Omega _{2}^{\ast }\Omega _{3}^{\ast
}-[i\left( \delta -\omega _{12}\right) +\frac{\Gamma _{2}}{2}]\Omega
_{1}^{\ast }\Omega _{4}\right]  \notag \\
& +i\frac{\mathcal{A}_{2}\left( 0\right) }{\mathcal{D}\left( \delta -\omega
_{12}\right) }\left[ i\left( \delta -\omega _{12}\right) [i\left( \delta
-\omega _{12}\right) +\frac{\Gamma _{1}}{2}]\Omega _{3}^{\ast }-\Omega
_{1}^{\ast }\Omega _{2}\Omega _{4}+\Omega _{1}^{\ast }\Omega _{1}\Omega
_{3}^{\ast }\right]  \notag \\
& -\frac{\mathcal{A}_{3}\left( 0\right) }{\mathcal{D}\left( \delta -\omega
_{12}\right) }[i\left( \delta -\omega _{12}\right) [i\left( \delta -\omega
_{12}\right) +\frac{\Gamma _{1}}{2}][i\left( \delta -\omega _{12}\right) +%
\frac{\Gamma _{2}}{2}]  \notag \\
& +[i\left( \delta -\omega _{12}\right) +\frac{\Gamma _{2}}{2}]\left\vert
\Omega _{1}\right\vert ^{2}+i\left( \delta -\omega _{12}\right) \left\vert
\Omega _{2}\right\vert ^{2}]  \notag \\
& +i\frac{\mathcal{B}\left( 0\right) }{\mathcal{D}\left( \delta -\omega
_{12}\right) }\left[ -\Omega _{1}\Omega _{2}^{\ast }\Omega _{3}^{\ast }-%
\left[ i\left( \delta -\omega _{12}\right) +\frac{\Gamma _{1}}{2}\right] %
\left[ i\left( \delta -\omega _{12}\right) +\frac{\Gamma _{2}}{2}\right]
\Omega _{4}-\Omega _{4}\left\vert \Omega _{2}\right\vert ^{2}\right] .
\end{align}%
Herein%
\begin{align}
\mathcal{D}\left( \delta \pm \omega _{12}\right) ,(\delta )& =\left( \delta
\pm \omega _{12}\right) ^{4},\delta ^{4}-i(\frac{\Gamma _{1}}{2}+\frac{%
\Gamma _{2}}{2}+\frac{\Gamma _{3}}{2})\left( \delta \pm \omega _{12}\right)
^{3},\delta ^{3}  \notag \\
& -(\frac{\Gamma _{1}\Gamma _{3}}{4}+\frac{\Gamma _{2}\Gamma _{3}}{4}+\frac{%
\Gamma _{1}\Gamma _{2}}{4}+\sum_{i=1}^{4}\left\vert \Omega _{i}\right\vert
^{2})\left( \delta \pm \omega _{12}\right) ^{2},\delta ^{2}  \notag \\
& +i[\frac{\Gamma _{1}\Gamma _{2}\Gamma _{3}}{8}+(\frac{\Gamma _{1}}{2}+%
\frac{\Gamma _{2}}{2})\left\vert \Omega _{4}\right\vert ^{2}+\frac{\Gamma
_{1}}{2}\left\vert \Omega _{3}\right\vert ^{2}+\frac{\Gamma _{3}}{2}%
\left\vert \Omega _{2}\right\vert ^{2}  \notag \\
& +(\frac{\Gamma _{2}}{2}+\frac{\Gamma _{3}}{2})\left\vert \Omega
_{1}\right\vert ^{2}]\left( \delta \pm \omega _{12}\right) ,\delta +(\frac{%
\Gamma _{1}\Gamma _{2}}{4}\left\vert \Omega _{4}\right\vert ^{2}+\frac{%
\Gamma _{2}\Gamma _{3}}{4}\left\vert \Omega _{1}\right\vert ^{2}  \notag \\
& +\left\vert \Omega _{2}\right\vert ^{2}\left\vert \Omega _{4}\right\vert
^{2}+\left\vert \Omega _{1}\right\vert ^{2}\left\vert \Omega _{3}\right\vert
^{2}-\Omega _{1}^{\ast }\Omega _{2}\Omega _{3}\Omega _{4}+\Omega _{1}\Omega
_{2}^{\ast }\Omega _{3}^{\ast }\Omega _{4}^{\ast }).
\end{align}%
The spontaneous emission spectrum for the atom initially in
$\left\vert
B\right\rangle $ can then be calculated analytically from $%
\mathbb{S(\delta )}=\Gamma _{n}\left\vert C_{k}\left(
t\longrightarrow \infty \right) \right\vert ^{2}/2\pi \left\vert
g_{k}^{(n)}\right\vert ^{2},$ $\left( n=1-3\right) $. Further, to
interpret the result we can write  $C_{k}\left( t\rightarrow
\infty
\right) $ as \cite%
{F-Ghafoor}

\begin{align}
C_{k}\left( t\rightarrow \infty \right) & \varpropto \sum_{i=1}^{4}\frac{%
g_{k}^{\ast }}{\boldsymbol{\digamma }_{1}}\left[ \frac{\left[ \Omega
_{2}\Omega _{3}\Omega _{4}-\left[ i\left( \lambda _{i}+\omega _{12}\right) +%
\frac{\Gamma _{3}}{2}\right] \left[ i\left( \lambda _{i}+\omega _{12}\right)
+\frac{\Gamma _{2}}{2}\right] \Omega _{1}-\Omega _{1}\left\vert \Omega
_{3}\right\vert ^{2}\right] \Bbbk _{i}}{\Delta -\lambda _{i}}\right]  \notag
\\
& +\sum_{i=1}^{4}\frac{g_{k}^{\ast }}{\boldsymbol{\digamma }_{2}}\left[
\frac{\left[ \left( i\mu _{i}+\frac{\Gamma _{1}}{2}\right) \Omega _{2}\Omega
_{3}+\left( i\mu _{i}+\frac{\Gamma _{3}}{2}\right) \Omega _{1}\Omega
_{2}^{\ast }\right]
\mathbb{R}
_{i}}{\Delta -\mu _{i}}\right]  \notag \\
& +\sum_{i=1}^{4}\frac{g_{k}^{\ast }}{\boldsymbol{\digamma }_{3}}\left[
\frac{\left[ -\Omega _{1}\Omega _{2}^{\ast }\Omega _{3}^{\ast }-\left[
i\left( \kappa _{i}-\omega _{12}\right) +\frac{\Gamma _{1}}{2}\right] \left[
i\left( \kappa _{i}-\omega _{12}\right) +\frac{\Gamma _{2}}{2}\right] \Omega
_{4}-\Omega _{4}\left\vert \Omega _{2}\right\vert ^{2}\right]
\mathbb{C}
_{i}}{\Delta -\kappa _{i}}\right]
\end{align}%
where, $\boldsymbol{\digamma }_{j}$ $\left( j=1-3\right)
=R_{1}^{3}(R_{3}-R_{4})(R_{2}^{2}+R_{3}R_{4}-2R_{2}\lambda
_{4})+R_{2}^{3}[(R_{4}-R_{3})(R_{1}^{2}+R_{3}R_{4})-R_{1}(R_{4}^{2}-R_{3}^{2})]+R_{3}^{3}(R_{4}-R_{1})[R_{1}R_{4}-R_{2}(R_{4}-R_{2})]+R_{4}^{3}[(R_{3}-R_{1})(R_{1}^{2}-R_{1}R_{2}-R_{2}^{2})+R_{2}(R_{4}^{2}-R_{1}^{2})]+R_{1}R_{2}R_{3}R_{4}[R_{1}(1+R_{4})-2R_{3}R_{4}],
$ with $R^{\prime }s$ $\Longrightarrow \lambda _{i},$ $\mu _{i},$ $\kappa
_{i} $ for each term being the roots of quartet Eqs. (9), respectively. Also
$M_{j}\left( 1-3\right) =\sum\limits_{i=1}^{4}\Bbbk _{i},%
\mathbb{R}
_{i},%
\mathbb{C}
_{i}=(R_{2}^{2}+R_{3}R_{4})(R_{4}-R_{3})-R_{2}(R_{4}^{2}-R_{3}^{2}),$ $%
(R_{3}^{2}+R_{4}R_{1})(R_{1}-R_{4})-R_{3}(R_{1}^{2}-R_{4}^{2}),$ $%
(R_{4}^{2}+R_{1}R_{2})(R_{2}-R_{1})-R_{4}(R_{2}^{2}-R_{1}^{2}),$ $%
(R_{1}^{2}+R_{2}R_{3})(R_{3}-R_{2})-R_{1}(R_{3}^{2}-R_{2}^{2}).$ Further,
the phases associated with the two microwave fields are $\Omega _{2}=$ $%
\left\vert \Omega _{2}\right\vert e^{i\varphi _{2}}$ and $\Omega
_{3}=\left\vert \Omega _{3}\right\vert e^{i\varphi _{3}}$ while $\Omega
_{1}=\left\vert \Omega _{1}\right\vert $ and $\Omega _{4}=\left\vert \Omega
_{4}\right\vert $ are real. The spontaneous emission spectrum $\mathbb{%
S(\delta )}$ for any values of spectroscopic parameters is then given by%
\begin{equation}
\mathbb{S(\delta )}=\Gamma _{1}\left\vert \sum_{i=1}^{4}\frac{\chi
_{i}+i\tau _{i}}{\left( \delta -\varrho _{i}\right) +i\sigma _{i}}%
\right\vert ^{2}+\Gamma _{2}\left\vert \sum_{i=1}^{4}\frac{\gamma
_{i}+i\varpi _{i}}{\left( \delta -\sigma_{i}\right) +i\theta _{i}}%
\right\vert ^{2}+\Gamma _{3}\left\vert \sum_{i=1}^{4}\frac{\epsilon
_{i}+i\varepsilon _{i}}{\left( \delta -\eta _{i}\right) +i\rho _{i}}%
\right\vert ^{2},
\end{equation}%
where all the symbols appear for appropriate integers associated
with a chosen set of spectroscopic parameters of the system. Now,
Eq. (11) consists of three parts where every ones is associated
with four dressed-states. Here we neglected the interference terms
among the three sets of dressed-states due to large separation
among the bare-state. Therefore, the spectrum consists, in general
of twelve peaks located at $\delta =\rho _{i}$, $\sigma _{i}$ and
$\theta _{i}$ with the peak heights ($\chi _{i}^{2}+\tau
_{i}^{2})/\varrho _{i}^{2},$ ($\gamma _{i}^{2}+\varpi
_{i}^{2})/\sigma_{i}^{2}$ and ($\epsilon _{i}^{2}+\varepsilon
_{i}^{2})/\eta _{i}^{2}$ (for $i=1-4$), respectively.

Next, I examine the condition for a trapping state in this system
for SGC and set the constant part of the characteristics equation
to zero. The resulting trapping condition when
satisfy the equation, $%
(\Gamma _{1}\Gamma _{2}/2+\Gamma _{2}\Gamma _{3}/2+4\left\vert
\Omega \right\vert )-i\left( 4\left\vert \Omega \right\vert
^{2}\sin \varphi _{2}\right) =0,$ where
$\sum\limits_{i=1}^{4}\left\vert \Omega _{i}\right\vert
=\left\vert \Omega \right\vert$. In this equation the imaginary
part can be zero if $\varphi _{2}=0$, while the vanishing of the
real part requires the un-physical condition of negative decay
rates. Therefore, there is no trapped dressed state due to SGC. In
principle, the physics is different in this system, and it is
based on FGC where the quantum coherence is generated by the
combinational effect of phases of the two microwave driven fields
and the amplitudes of all the driving fields. To get the trapping
condition, we set the numerator of the central major part of the
spectrum equation to zero i.e., $$ i\delta \left( \left\vert
\Omega _{3}\right\vert \left\vert \Omega _{4}\right\vert
e^{i\varphi _{3}}+\left\vert \Omega _{1}\right\vert \left\vert
\Omega _{2}\right\vert e^{-i\varphi _{2}}\right) +(\frac{\Gamma
_{1}}{2}\left\vert \Omega _{3}\right\vert \left\vert \Omega
_{4}\right\vert e^{i\varphi _{3}}+\frac{\Gamma _{3}}{2}\left\vert
\Omega _{1}\right\vert \left\vert \Omega _{2}\right\vert
e^{-i\varphi _{2}})=0.$$ Obviously the first part can be zero when
the phases, $\varphi _{2}\left(
\varphi _{3}\right) =\pi \left( 0\right) ,0\left( \pi \right) $ and $%
\left\vert \Omega _{3}\right\vert \left\vert \Omega
_{4}\right\vert =\left\vert \Omega _{1}\right\vert \left\vert
\Omega _{2}\right\vert $ while the vanishing of the second part
requires $\Gamma _{1}=\Gamma _{3}.$ Remarkably, these conditions
which are novel and externally controllable unlike the ones in
Refs. \cite{Zhu-Scully,Paspalakis}. The second major result is the
simultaneously cancellation of the four spectral lines arising
from the central decaying \emph{bare-energy-state} unlike the one
spectral line of the unique dressed state of the early studies.

Interestingly, if we extend to a system of simple loss based on
the Zeeman hyperfine Sodium D1 line with four ground states and
one excited decaying
states driven by two microwave and two optical fields [see Fig. 1(b)] \cite%
{GhafoorF}. In this system the whole brightened atom can be
darkened under its trapping condition, $\left\vert \Omega
_{o_{1}}\right\vert \left\vert \Omega _{m_{1}}\right\vert
e^{i\varphi _{3}}+\left\vert \Omega _{m_{2}}\right\vert \left\vert
\Omega _{o_{2}}\right\vert e^{-i\varphi
_{2}}=0.$ In getting this condition we assume $\Omega _{o_{1,2}}=$ $%
\left\vert \Omega _{o_{1,2}}\right\vert e^{i\varphi _{3,2}}$ and
$\Omega _{m_{1,2}}=\left\vert \Omega _{m_{1,2}}\right\vert .$ The
spontaneous emission spectrum is calculated from $G_{k}\left(
t\rightarrow \infty \right) =i\delta \left( \left\vert \Omega
_{o_{1}}\right\vert \left\vert \Omega _{m_{1}}\right\vert
e^{i\varphi _{3}}+\left\vert \Omega _{m_{2}}\right\vert \left\vert
\Omega _{o_{2}}\right\vert e^{-i\varphi _{2}}\right)
/\mathcal{D}\delta ),$ ( if in $\mathcal{D}\left( \delta \right) $
of Eq. (9) $\Gamma _{1}=\Gamma _{2}=0$ and $\Gamma _{3}=\Gamma $).
This simplified version can also be realized in a laboratory if we
select the four ground states of D1 lines i.e., $\left\vert
3S_{1/2},F_{2}=2,m_{F}=1\right\rangle (\left\vert g_{1}\right\rangle ),$ $%
\left\vert 3S_{1/2},F_{2}=2,m_{F}=0\right\rangle (\left\vert
g_{2}\right\rangle ),$ $\left\vert
3S_{1/2},F_{1}=1,m_{F}=1\right\rangle (\left\vert
g_{3}\right\rangle )$ and one excited state $\left\vert
3P_{1/2},F_{1}=1,m_{F}=0\right\rangle (\left\vert e\right\rangle
).$ The states $\left\vert 3S_{1/2},F_{2}=2,m_{F}=1\right\rangle
(\left\vert g_{1}\right\rangle )$ and $\left\vert
3S_{1/2},F_{2}=2,m_{F}=0\right\rangle (\left\vert
g_{2}\right\rangle )$ are coupled with the state $\left\vert
3S_{1/2},F_{1}=1,m_{F}=1\right\rangle (\left\vert
g_{3}\right\rangle )$ by
two microwave fields while they are coupled with the excited decaying state $%
\left\vert 3P_{1/2},F_{1}=1,m_{F}=0\right\rangle (\left\vert
e\right\rangle ) $ by two optical fields. The linkage of the
excited state is considered with the fourth ground state,
$\left\vert 3S_{1/2},F_{1}=1,m_{F}=0\right\rangle (\left\vert
g\right\rangle )$ via vacuum field modes.

\begin{figure}[t]
\centering
\includegraphics[width=4in]{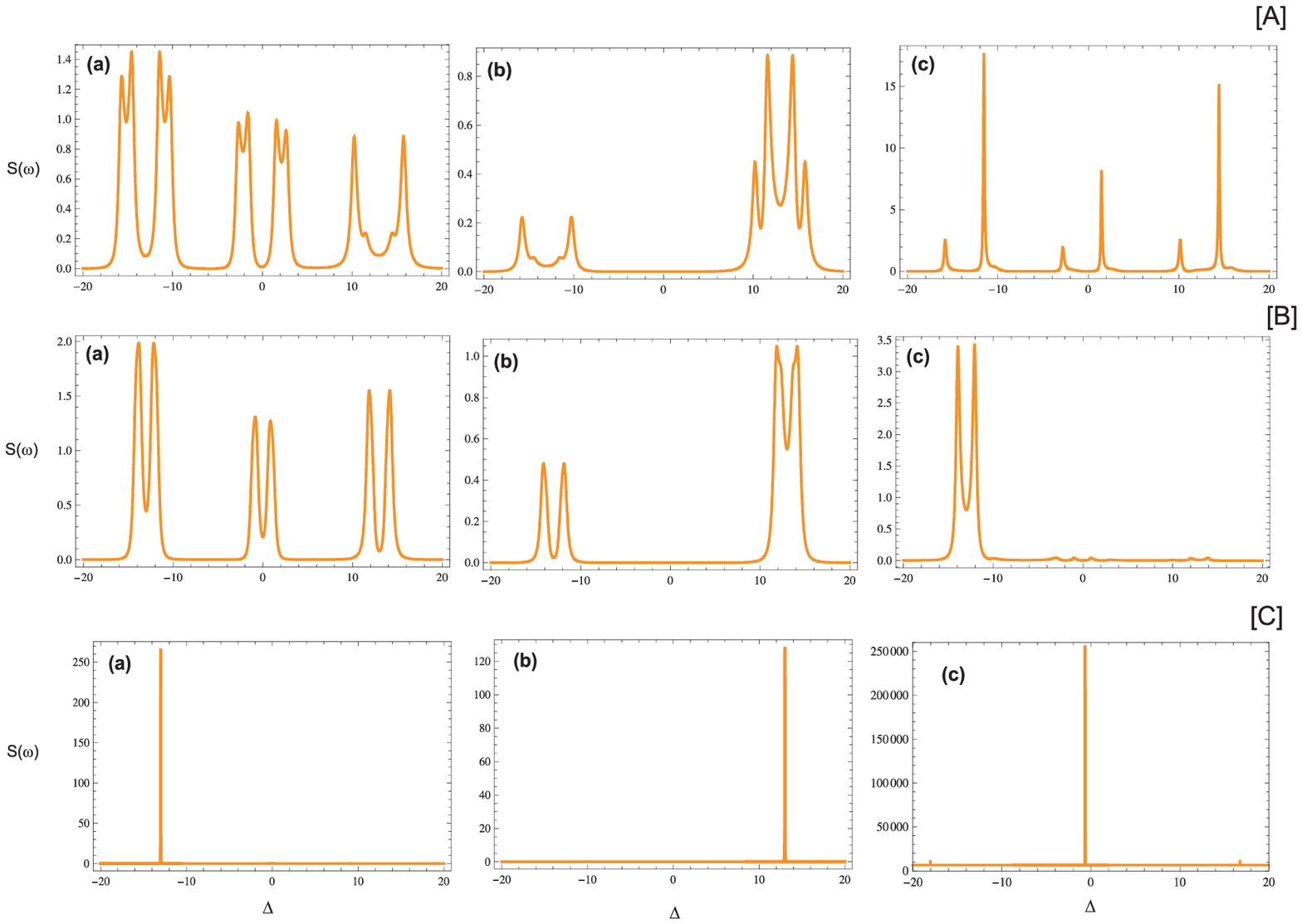}
\caption { [A]: Here $\Gamma _{1,2,3}=\Gamma $ and $\protect\omega
_{12}=13\Gamma
$. $\mathbb{%
S(\protect\delta )}$ (in unit of $\Gamma ^{-1}$) for $\left\vert
\Omega _{2,3}\right\vert=\Gamma,$ $\left\vert \Omega
_{1,4}\right\vert $ $=2\Gamma, $ for $\left\vert \Omega _{1,3}\right\vert$=$0.5\Gamma ,$ $\left\vert \Omega _{2,4}\right\vert $ $%
=0.9\Gamma,$ and for$\left\vert \Omega _{1,3}\right\vert=0.5\Gamma ,\left\vert \Omega _{2,4}\right\vert$ $%
=0.9\Gamma $ the values of $\protect\varphi _{2}\left(
\protect\varphi _{3}\right)$ for each case are (a) $0\left(
\protect\pi \right) $ (b) $\protect\pi \left( 0\right)$ (c)
$\protect\pi /2\left( 3\protect\pi /2\right).$} \label{figure3}
\end{figure}

Generally, inspecting the analytical expression for the
spontaneous emission spectrum in limiting cases for the scheme of
Fig. 1(a), we predicted the spectrum of a decaying of two-level
atom \cite{M. O. Scully}, of the scheme
of Autler-Towenes doublet \cite{Autler}, of the scheme of Paspalakis \emph{%
et al.} \cite{Pasp}, of the scheme of quantum beat laser \cite{M. O. Scully}%
, and of the scheme of Autler-Townes quartuplet spectroscopy \cite{GhafoorF}%
. Further, the analysis of Eq. (11) agrees well with the plot of
the analytical results of this system displaying twelve peaks
spectrum [see Fig. 2[A]], where each four are associated with the
dressed-state of the three bare-state. However, under the trapping
condition the four peaks originating from the central bare-state
are completely cancelled, while the side two sets of dressed-state
contribute significantly with enhanced values for the one set over
the other. The phase effect for all the fields is similar.
Therefore, keeping one symmetric of the other for the microwave
fields results in compensation of their atomic population transfer
under the trapping condition. In this way, the two symmetric
phases prevent the atom from decaying from the four dressed-state
of the central bare-energy-state leaving it completely darken.
Almost $41\%$ of the population is trapped in the excited state in
this case. Further, varying only the two phases individually from
$\pi $ to $\pi /2(3\pi/2)$ we get maximum narrowing for the two
central peaks while there is population transfer to the next
dressed state if the phases is varied further symmetrically (not
shown).

\begin{figure}[t]
\centering
\includegraphics[width=4in]{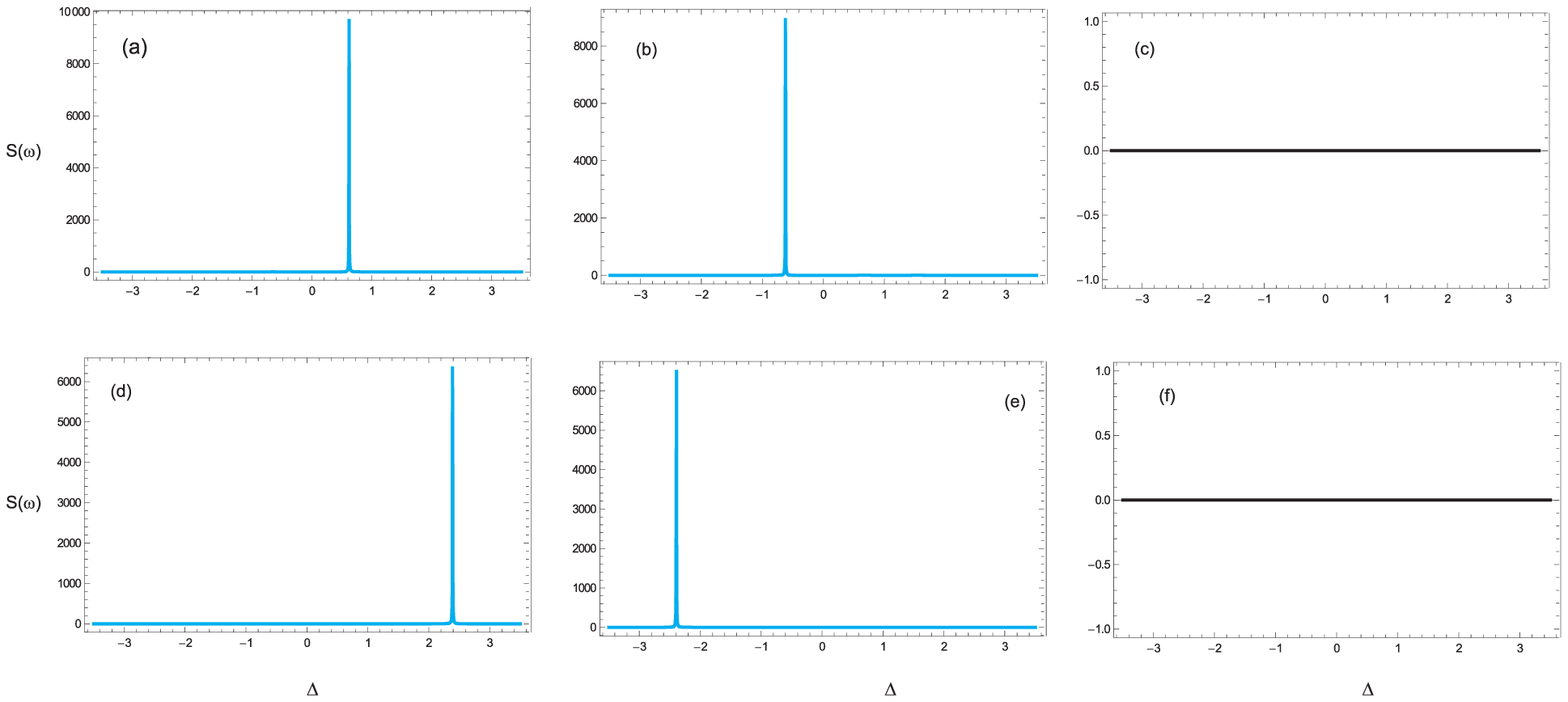}
\caption{$\mathbb{S(\delta )}$ (in unit of $\Gamma ^{-1}$) for
\textbf{(a)[(b)]} $\varphi _{2}\left( \varphi _{3}\right) =\pi
/2\left( 3\pi /2\right)[3\pi /2\left( \pi /2\right) ] $,
$\left\vert \Omega _{o_{2},o_{1}}\right\vert =0.5\Gamma $ and
$\left\vert \Omega _{m_{2},m_{1}}\right\vert =\Gamma .$
\textbf{(d)[(e)]} $\varphi _{2}\left( \varphi _{3}\right) =3\pi
/2\left( \pi /2\right)[\pi /2\left( 3\pi /2\right) ]$, $\left\vert
\Omega _{m_{2},m_{1},o_{2}}\right\vert =\Gamma ,$ and $\left\vert
\Omega _{o_{1}}\right\vert =0.1\Gamma \lbrack$ \textbf{(c)[(f)]} for examples, $%
\varphi _{2}\left( \varphi _{3}\right) =\pi \left( 0\right)$ and
$\left\vert \Omega _{m_{2},m_{1}o_{2},o_{1}}\right\vert =1\Gamma\
[\left\vert \Omega _{m_{2},m_{1}}\right\vert =1\Gamma, \left\vert
\Omega _{o_{2}o_{1}}\right\vert =2\Gamma].$
 } \label{figure5}
\end{figure}

Remarkably, with some appropriate relative strengths, the central
two peaks of the three decaying bare-state suppress extremely
while enhancing the sides one accordingly. However, when $\varphi
_{2}\left( \varphi _{3}\right) \longrightarrow \pi (0)$ the
trapping condition is satisfied, the two enhanced spectral lines
of the central bare-state is cancelled [see Fig. 2[B](b)] reducing
the area under the curve by $34\%$. However, no trapping state is
there when $\varphi _{2}\left( \varphi _{3}\right) \longrightarrow
\pi (0)$ [see Fig. 2[B](c)] except narrowing the spectral spectral
lines. Moreover, there is a variety of very narrow single-peaked
spectra for at least three different locations even with the one
satisfying the trapping condition [see Fig. 2[C](b)] and compare
its area under the curve with Fig. 2[C](a)]. This allows effective
storage and manipulation of our one-atom quantum system like the
two-atom quantum system of Ref. \cite{2-atom} but with the
advantage of $47\%$ population trapping in the upper excited
bare-state. Of course, this FGC-based result is novel and
remarkable as compared with earlier related results.

Intriguingly, extending to a system of simple probability loss
[see Fig. 1(b)] which generally has four-peak spectral profile can
be manipulated to extremely narrowed-one-peak spectral profile at
different locations for different choices of phases and fields
strength [see Fig. 3(a-f)]. However, under the trapping condition
of this system, the only decaying dressed \emph{bare-energy-state}
can also be completely darkened due to FGC. This is a major result
meaning $100\%$ population trapping, a novel state of a darken
atom. The atom remains in the dark state until the trapping
condition is held on.

In conclusion, the FGC based atomic scheme is presented for
spontaneous emission cancellation in contrast to the traditional
SGC. The phases and strengths of the driven fields collectively
modify the spontaneous emission spectrum due to which one to
twelve peaks of varying widths arise. Further, experimentally easy
controllable trapping condition is explored for spontaneous
emission cancellation. This cancellation is from the whole set of
four dressed states associated with the \emph{central
bare-energy-state} of the three set of closely spaced hyperfine
decaying bare states. Extending this concept to a system of a
simple loss, based on real atomic system, the brightened atom can
also be darkened under its trapping condition, an interesting and
viably novel phenomenon. The control of phases of the driving
fields \cite{Phase} and the coupling of multiple fields with an
atomic system \cite{Zanch} are now laboratory realities. These may
be helpful in demonstrating the mechanism of the physical
phenomenon of FGC in a laboratory for the spontaneous emission
cancellation.


\begin{thebibliography}{99}
\bibitem{Agarwal} G. S. Agarwal, Quantum Optics (Springer-Verlag, Berlin,
1974).

\bibitem{Alzetta} G. Alzetta et al., Nuovo Cimento Soc. Ital. Fis. 36B, 5
(1974); E. Arimondo, in Progress in Optics, edited by E. Wolf (Elsevier,
Amsterdam, 1996), Vol. XXXV, p. 257.

\bibitem{Gray} P. L. Knight, J. Phys. B 12, 3297 (1979); D. A. Cardimona et al., J. Phys. B 15, 55 (1982);
D. Agassi, Phys. Rev. A 30, 2449 (1984).

\bibitem{Garraway} B. M. Garraway and P. L. Knight, Phys. Rev. A \textbf{54}%
,2379 (1969).

\bibitem{Zhu} S.-Y. Zhu, H. Chen, and H. Huang, Phys. Rev. Lett. \textbf{79,
}205 (1997); Lewenstein \emph{et al.} Phys. Rev. A \textbf{38},
808 (1988); S. Bay, P. Lambropoulos, and K. M$\o $lmer, Phys. Rev.
A \textbf{79,} 2654 (1997); A. G. Kofman, G. Kurizki, and B.
Sherman, J. Mod. Opt. \textbf{41, }353 (1994).

\bibitem{Kochar} For examples see, O. Kocharovskaya and Ya. I. Khanian, JETP Lett. \textbf{48}%
,630 (1988); O. Kocharovskaya and P. Mandel, Phys. Rev. A
\textbf{42}, 523 (1990).

\bibitem{Scully} M. O. Scully, Phys. Rev. Lett. \textbf{55}, 2802 (1975); W.
Schleich and M. O. Scully, Phys. Rev. A \textbf{37, }1261\textbf{\ }(1987);
J. Bergou, M. Orszag, and M. O. Scully, Phys. Rev. A \textbf{38, }754 (1988).

\bibitem{Paspalakis} E. Paspalakis and P. L. Knight, Phys. Rev. Lett.
\textbf{81, } 293 (1998).

\bibitem{Zhu-Scully} S.-Y. Zhu and M. O. Scully, Phys. Rev. Lett. \textbf{76,%
} 388 (1996).

\bibitem{2-atom} E. M. Macove and C. H. Keitel, Phys. Rev. Lett. \textbf{91, }
123601 (2003).

\bibitem{F-Ghafoor} No trapping state exists in "F. Ghafoor \emph{et al.} Phys. Rev. A \textbf{62}, 13811 (2000)". In Ref.
"J. H. Wu et al., Phys. Rev. A \textbf{74}, 033816 (2006)" Fano
type profile exists and the spectrum is like the Autler-Townes
triplet \cite{GhafoorF} and obviously there is no CPT.

\bibitem{Hwang-Zhu} Lee \emph{et al.}  Phys. Rev. A \textbf{55}, 4454 (1997).

\bibitem{Hakuta} Hakuta \emph{et al.} Phys. Rev. Lett.
\textbf{66}, 596 (1991).

\bibitem{Xia} H.-R. Xia, C.-Y. Ye, and S.-Y. Zhu, Phys. Rev. Lett. \textbf{77%
}, 1032 (1996)\nolinebreak.

\bibitem{Li} Li Li, X. Wang, J. Yang, G. Lazarov, J. Qi, A. M. Lyyra, Phys.
Rev. Lett. \textbf{84}, 4016 (2000).

\bibitem{Ghafoor00} F. Ghafoor, Phys. Rev. A \textbf{84}, 063849 (2011).

\bibitem{Herz} G. Herzberg, \textit{Molecular Spectra and Molecular
Structure I; Spectra of Diatomic Molecules }(Van Nostrand, Princeton, 1950)

\bibitem{Wang} See for example, Z. G. Wang and H. R. Xia, \textit{Molecular
and Laser spectroscopy} (Springer-Verlag, Berlin, 1991).\nolinebreak

\bibitem{GhafoorF} F. Ghafoor, Phy. Rev. A (revision); \ For Autler-Townes
triplet spectrum see, F. Ghafoor, Opt. Commun. \textbf{284},1913 (2011); F.
Ghafoor, S. Qamar, S.- Y. Zhu, and M. S. Zubairy, Opt. Commun. \textbf{273},
464 (2007).

\bibitem{Zanch} For example, see Zuo \emph{et al}. Phys. Rev. Lett. 97, 193904 (2004).

\bibitem{M. O. Scully} M. O. Scully and M. S. Zubairy, \textit{Quantum Optics%
}\textrm{\ }(Cambridge University Press, Cambridge, 1997).

\bibitem{Autler} H. Autler and C. H. Townes, Phys. Rev. A \textbf{100}, 703
(1955); P. L. Knight and P. W. Milonni, Phys. Rep. \textbf{66}, 23 (1980).

\bibitem{Pasp} E. Paspalakis, C. H. Keitel, and P. L. Kinght, Phys. Rev. A,
58, 4868 (1998).

\bibitem{Phase} L. Zhu et al., Science 270, 77 (1995); C. Chen and D. S.
Elliott, Phys. Rev. Lett. 65, 1737 (1990). \newpage
\end{thebibliography}
\end{document}